FRONT MATTER

# Boron-assisted synthesis of compositionally complex amorphous oxides via short-range-order-constrained generative design

Honglin Li[1, 2, †], Chuhao Liu[3, 4, *], Yongfeng Guo[2, †], Xiaoshan Luo[1, †], Yijie Chen[5, †], Guangsheng Liu[6], Yu Li[2], Zhenyu Wang[1], Jianzhuo Wu[2], Shouwei Zuo[7], Zhen Luo[3], Cheng Peng[3], Qinyu Jiang[4], Jialu Li[6], Cheng Ma[1], Zhuohang Xie[1], Jian Lv[1], Yufei Ding[8], Huabin Zhang[7], Mufan Li[3, *], Yanchao Wang[1, *] and Wan-Lu Li[2, 6, *]

[1]*Key Laboratory of Material Simulation Methods and Software of Ministry of Education, College of Physics, Jilin University, Changchun, China*
[2]*Aiiso Yufeng Li Family Department of Chemical and Nano Engineering, University of California, San Diego, La Jolla, CA, USA*
[3]*College of Chemistry and Molecular Engineering, Peking University, Beijing, China*
[4]*Institute of Molecular Engineering Plus, College of Chemistry, Fuzhou University, Fuzhou, China*
[5]*Institute of Modern Physics, Fudan University, Shanghai, China*
[6]*Program in Materials Science and Engineering, University of California, San Diego, La Jolla, CA, USA*
[7]*Center for Renewable Energy and Storage Technologies, Physical Science and Engineering Division, King Abdullah University of Science and Technology, Thuwal, Kingdom of Saudi Arabia*
[8]*Department of Computer Science and Engineering, University of California, San Diego, La Jolla, CA, USA*

[*]Corresponding author. Email:
wal019@ucsd.edu, wyc@calypso.cn, mufanli@pku.edu.cn, chuhao_liu@fzu.edu.cn

[†]These authors contributed equally to this work.

## Abstract

Engineering short-range atomic order in amorphous materials offers a promising yet still underexplored route to high-performance solids. Here, we establish a boron-assisted amorphization strategy through ApolloX, a theory-guided, short-range-order-constrained generative framework for identifying low-energy amorphous configurations in compositionally complex multimetal $BO_x$ systems. Using $FeCoNiMoBO_x$ as a representative model platform, ApolloX predicts an ensemble of candidate amorphous configurations across systematically varied boron contents. *Ab initio* molecular dynamics simulations based on these configurations show that increasing boron content suppresses atomic diffusion and disfavors crystallization, with the stabilization of $BO_3$-centered local motifs emerging as a key structural feature associated with enhanced amorphization propensity. Guided by these predictions, we synthesize three representative $FeCoNiMoBO_x$ compositions with distinct boron contents and use synchrotron-based scattering and electron microscopy to verify compositional fidelity, structural homogeneity, and the targeted amorphous features, thereby experimentally validating the boron-regulated structural evolution predicted by theory. Beyond this representative system, the same strategy is extended to a broader library of multimetal $BO_x$ amorphous compositions spanning diverse metal combinations and boron loadings, demonstrating that the boron-assisted route is not limited to a single $FeCoNiMoBO_x$ family but is broadly transferable across compositionally complex amorphous oxides. Overall, our results establish boron incorporation as a practical design variable for tuning short-range order and amorphization in multicomponent oxides, and provide a general framework for the theory-guided discovery of compositionally complex amorphous materials with tunable properties.

## Introduction

Amorphous materials play an important role across ceramics, glasses, catalysts, and energy-storage systems. The high compositional tunability makes them an excellent platform for multi-component solid synthesis at the atomic-level and has led to their extensive applications due to their diverse local coordination environments and cocktail effect[1–3]. Unlike crystalline solids, which possess long-range periodicity, amorphous phases lack long-range order and are instead dominated by short-range atomic order (SRO)[4–7]. Consequently, variations in local bonding motifs and coordination statistics often exert a stronger influence on functional properties than the overall (global) symmetry. Recent experimental studies have demonstrated that differences in SRO can modulate electrochemical behavior, including oxygen evolution, oxygen reduction, and redox activity in battery cathodes[8–11]. These findings highlight the critical importance of understanding and tuning SRO in amorphous systems. Yet, establishing quantitative links between composition, local structure, and amorphization behavior in multicomponent solids remains highly challenging due to the unpredictable nature of the amorphous structure.

Amorphous solids exhibit a rugged energy landscape in which many metastable configurations are accessible even at a fixed global composition[12], and the specific SRO motifs realized experimentally often depend sensitively on the synthesis pathway[13]. In practice, melt–quench molecular dynamics (MD) is highly effective for modeling amorphous structures formed under fast annealing conditions, capturing key trends of amorphization and local structural evolution[14–16]. However, directly simulating experimentally relevant slow synthesis rates with MD is often computationally prohibitive[17–21]. Therefore, identifying low-energy local environments that are accessible under slow synthesis conditions can serve as a practical surrogate for direct slow-rate MD. In 1953, Brown et al. first reported the preparation of amorphous catalysts using boron-based chemistry, thus opening up the application of amorphous materials in catalysis[22]. Our previous work also confirmed that B doping can effectively reduce the coordination number of metal sites in metal oxides[23], but the associated boron-centered structural changes could not be fully resolved. Extending these insights to chemically complex, multimetal oxides requires a framework that can systematically explore how composition governs diffusion behavior, local bonding motifs, and the resulting amorphous structure.

In this study, we investigate how boron content regulates SRO and amorphization in multimetal $BO_x$ systems by integrating a boron-assisted synthesis route with ApolloX, a generative modeling framework that samples compositionally consistent local environments under prescribed SRO constraints. The model system comprises OER-relevant transition metals (Fe, Co, Ni, and Mo)[24–26] combined with boron oxide. As a textbook glass former[27], $B_2O_3$ features a small ionic radius and can readily integrate into metal–oxygen networks, promoting amorphization by disrupting long-range periodicity and suppressing phase separation, crystalline domains, and interfacial boundaries[28]. Using $FeCoNiMoBO_x$ as a representative platform, we combined ApolloX with *ab* initio molecular dynamics (AIMD) to quantify how boron concentration modulates $BO_3$ formation, metal–oxygen coordination motifs, and atomic diffusion—key factors that collectively control amorphization and can ultimately impact oxygen evolution activity. Guided by these predictions, we first synthesize three representative compositional groups and perform rigorous, multimodal characterization to confirm compositional fidelity, structural homogeneity, and the targeted amorphous/SRO features, thereby supporting the feasibility of the proposed synthesis strategy and the boron-regulated structural trends predicted by theory. Building on this validation, we further extend the approach to a broad library of distinct multimetal $BO_x$ amorphous compositions (spanning diverse metal combinations and boron loadings), demonstrating the generality and transferability of the boron-assisted route for composition-driven tuning of amorphous structures in multicomponent oxides.

# Results

The ApolloX-guided design framework for multi-element amorphous solids is schematically summarized in Fig. 1. In this workflow, ApolloX first generates statistically representative local short-range-ordered configurations for chemically complex multi-metal $BO_x$ systems (Fig. 1A), providing atomic-level insight into how different metal and boron contents reshape the local coordination environment. These configurations are then used as starting points for AIMD simulations, which reveal the dynamic formation of $BO_3$ motifs, the associated lower diffusion rates, and the resulting enhancement of amorphization. Guided by these theoretical insights, we design a composition-controlled synthesis strategy (Fig. 1B), in which both the identity or ratio of metal cations and the boron content are tuned so that higher boron loadings promote stronger amorphization. Finally, this integrated computational–experimental strategy culminates in synthesizing the predicted compositions and validating their amorphous structure and elemental distributions using electron microscopy and spectroscopy (Fig. 1C). Building on this framework, we next establish a boron-assisted amorphization platform for chemically complex multi-metal $BO_x$ catalysts.

## A. ApolloX: An SRO-constrained framework for predicting low-energy local environments in multi-element amorphous solids

For amorphous solids, the final structure is highly sensitive to the kinetic pathway of synthesis, with slower assembly generally yielding more relaxed, lower-energy local environments. To efficiently identify the low-energy structural motifs accessible to a specific multicomponent composition, we developed ApolloX (Automatic Prediction by generative mOdel for Large-scaLe Optimization of X-composition materials). Unlike traditional brute-force, atom-by-atom exploration of the full configurational space, ApolloX employs a SRO-constrained conditional generative framework. It predicts physically plausible, low-energy SRO patterns, represented as local coordination environments, that are consistent with prescribed thermodynamic conditions. This strategy transforms amorphous structure discovery into a guided optimization within a reduced SRO space, substantially improving search efficiency.

As outlined in Fig. 2A, the ApolloX workflow comprises four main stages. First, we employ the pair-density matrix (PDM) as a compact, size-independent $K$ x $K$ descriptor (where $K$ is the number of element types) to capture local chemical environments and SRO statistics (Fig. 2B) (Supplementary Section A). Second, an in-house conditional generative model (Cond-CDVAE, Fig. 2C) is trained on 10,000 reference structures to learn the statistical mapping between PDMs and atomic configurations. At inference, the model generates atomic structures conditioned on target PDMs. Third, candidates are rapidly relaxed and screened for thermodynamic stability—based on formation enthalpy and configurational entropy (Fig. 2D)—using the fine-tuned DPA-2 machine-learning interatomic potential. Finally, a population-based particle swarm optimization (PSO, managed via PySwarms, Fig. 2E) iteratively refines the SRO toward lower-energy configurational manifolds. To preserve structural diversity and prevent trapping in local minima, 60% of the lowest-energy structures are retained in each PSO cycle, while the remainder are newly generated.

To demonstrate the methodological advantage of ApolloX for compositionally complex disordered materials, we benchmarked it on a chemically diverse boron-rich oxide, $Fe_{12}Co_{12}Ni_{12}Mo_{12}B_{12}O_{60}$ (120 atoms; Supplementary Section B), chosen as a representative charge-balanced multicomponent amorphous system. Starting from 10,000 randomly generated configurations constructed by elemental substitution on a BCC parent phase, with analogous tests on FCC and HCP lattices (Supplementary Section C), ApolloX accurately reconstructed the target chemical short-range order (CSRO), yielding mean relative PDM errors typically below 20–25%. Benchmark comparisons with representative baseline approaches, including random initialization, special quasirandom structures (SQS), Monte Carlo (MC)-only sampling, and Metropolis Monte Carlo (MMC), further demonstrate its improved efficiency in identifying low-energy structures. Among 1,500 evaluated structures, 1.7% of ApolloX-generated configurations fell below an energy threshold of −960 eV, compared with 1.2% for the extensive MMC baseline and below 0.3% for the other methods.

Notably, ApolloX identified five ultralow-energy configurations below −990 eV, whereas such configurations were not observed by the baseline approaches. Beyond introducing PDM as a compact descriptor, the key innovation of ApolloX lies in reformulating amorphous-structure prediction as an efficient, directed search in a fixed-dimensional CSRO space. By projecting a large off-lattice amorphous configuration onto a compact $K \times K$ PDM representation and coupling this space with conditional generative modeling and particle-swarm optimization (PSO), ApolloX enables directed exploration of thermodynamically favorable amorphous motifs that would be difficult to access through direct optimization in the full atomic-coordinate space. Consistent with this design, the generated structures exhibit high coverage, uniqueness, and novelty across the configurational landscape (Supplementary Figs.

S1–S2 and Table S2), while PSO evolution from 100 generated seeds rapidly identifies low-energy amorphous configurations within only a few generations. Furthermore, analysis of 1,500 relaxed structures reveals a correlation between PDM deviation from the ground-state motif and the potential energy (Figs. S1e), confirming that the PDM-based representation captures the essential energetic information of the amorphous landscape and provides an effective optimization coordinate for low-energy amorphous structure discovery.

**B. Boron-controlled diffusion and amorphization tendency from MD**

To investigate how boron influences amorphization and local structure, we combine ApolloX with AIMD to examine a series of multimetal $BO_x$ compositions. ApolloX was used to generate SRO-consistent configurations containing 6%, 9%, and 12% boron (Group-1, -2, and -3, respectively) while keeping the relative metal ratios constant within a maximum deviation of  2%. These structures serve as starting points for assessing how incremental changes in boron concentration affect microstructure and stability. For each composition, 1,500 candidate structures are sampled and the lowest-energy configurations, which implicitly reflect both enthalpic and configurational entropy contribution, are selected for further analysis and to define the composition window later realized experimentally. The predicted local-structure maps and elemental enrichment heat maps (Fig. 3A) show that increasing boron content drives a transition from dispersed to aggregated metal distributions. Comparison of the atomic structures across the three groups reveals a clear evolution in SRO: in Group-1 (low boron), metal–O–metal connectivity dominates, whereas increasing boron causes oxygen to preferentially coordinate with boron, forming $BO_3$ motifs in which boron is threefold coordinated (Fig. 3B). At the highest boron level (Group-3), this shift toward $BO_3$ formation markedly reduces metal–oxygen interactions and promotes metal clustering (Fig. 3A). These trends highlight the pivotal role of boron in reshaping local SRO, driving a crossover from metal–O–metal connectivity toward B–O coordination and direct metal–metal interactions, consistent with experimental bonding analysis that indirectly confirms the presence of $BO_3$ units.

From a kinetic standpoint, crystallization requires local atomic rearrangements into ordered motifs, and its nucleation and growth rates are therefore set by atomic diffusion. When diffusion is slow, crystallization is hindered and the temperature–time window for forming or retaining an amorphous state expands[30, 31]. This diffusion–amorphization relationship is well established in solid-state amorphization and diffusion-couple systems: reduced atomic mobility (or slower structural relaxation) suppresses crystallization and promotes vitrification, thereby shaping the resulting amorphous structures[32–36]. In multicomponent and high-entropy systems, complex chemical environments further suppress diffusion and slow structural relaxation, enhancing amorphous stability[37, 38]. In such settings, the slowest diffusing species effectively acts as a kinetic bottleneck for structural restructuring[39–41] .To test this diffusion-controlled picture, we perform AIMD simulations with varying boron concentrations, heating each system from 300 K to 2050 K over 24 ps (Figs. 3C, D, F). Here the AIMD heating ramp is used as a comparative kinetic probe (relative mobility across compositions), rather than a direct model of experimental synthesis. Quantitative analysis of diffusion coefficients shows that the low-boron group exhibits a higher overall diffusivity (Supplementary Fig. S5). Fig. 3C reveals that increasing boron content systematically suppresses metal and boron atomic diffusion; among all elements, boron diffuses the slowest and its mobility decreases most steeply as its concentration increases. The time evolution of the mean squared displacement (MSD) for boron exhibits a pronounced stagnation plateau between 12.5 ps and 20 ps in the high-boron system, coincides with a transient "frozen" state. Structural snapshots (Fig. 3F) show that, during this plateau, boron atoms form locally stable $BO_3$ triangular motifs that percolate into a short-range network, strongly restricting boron mobility. Upon further heating, thermal energy disrupts the $BO_3$ framework, restoring diffusion.

Within this framework, boron-regulated short-range order, particularly the increasing prevalence of $BO_3$ -centered local environments together with the accompanying metal clustering, is associated with reduced atomic mobility and an enhanced tendency toward amorphization. This behavior suggests that the system can transiently access a low-mobility, short-range-ordered state before further thermal activation drives it toward a more disordered and diffusive regime. Figure 3D summarizes the proposed boron-assisted amorphization mechanism. Specifically, boron incorporation is proposed to introduce local geometric distortion through the formation of $BO_3$-centered motifs and boron-rich local environments, thereby disrupting long-range structural coherence and kinetically frustrating crystallization, which together favor the formation of an amorphous structure. As shown in Fig. 3e, the normalized B K-edge XANES spectra of Group-1, Group-2, and Group-3, together with commercial $B_2O_3$ as a reference, reveal a systematic evolution of the local boron coordination environment. The pronounced feature at ~194 eV is characteristic of trigonal $BO_3$ units, whereas the higher-energy shoulder at ~198 eV is assigned to tetrahedral $BO_4$ species. Notably, the $BO_3$-related feature increases progressively from Group-1 to Group-3, consistent with an increasing fraction of trigonal boron environments as the boron content increases.

To further validate our theoretical model of boron-assisted amorphization and confirm the robustness of the diffusion-based interpretation across a broader compositional space, we constructed an equal-ratio theoretical series containing 0–20% boron (g1–g11) (Supplementary Fig. S6). Using ApolloX to generate low energy candidates, we established a controlled framework to systematically quantify the macroscopic effects of boron. These candidate structures were subsequently evaluated through molecular dynamics simulations driven by machine-learning potential (MLP). Thermodynamically, formation-energy calculations reveal that the fundamental structural competition lies between local metal–O and B–O configurations (Supplementary Fig. S7). With the exception of molybdenum, the formation energy of $B_2O_3$ is comparable to or lower than those of the corresponding transition-metal oxides. This confirms that B–O motifs are highly competitive, locally redistributing oxygen from metal environments to borate-like configurations; this oxygen scavenging fragments the continuous metal–oxygen framework and drives the metal aggregation observed in our experiment-constrained maps (Fig. 3A). Kinetically, consistent with the diffusion-controlled mechanism discussed above, our MLP-driven simulations demonstrate a systematic decrease in both overall atomic diffusion (Supplementary Fig. S6a) and bond-orientational order parameters (Supplementary Fig. S6b) as the boron content increases.

Collectively, this broad compositional series confirms that boron promotes amorphization through a synergistic mechanism: thermodynamically, it introduces robust oxygen competition that disrupts long-range networks of metal oxides; kinetically, it acts as a universal slow-diffusing bottleneck that suppresses overall atomic mobility and stabilizes disordered short-range environments.

## C. Amorphous catalysts synthesis and experimental validation of boron-driven amorphization

Accompanied by the dynamical and thermodynamic insights provided by ApolloX, we developed a general boroxide-mediated route for the synthesis of amorphous multimetal oxide catalysts (Detail of the synthesis method as shown in Supplementary Section I). This protocol could be extended across a broad range of chemical complexities, including binary, ternary, quaternary, quinary, eight-component, ten-component and even more elements' systems (Fig. 4). Importantly, the XRD patterns of all synthesized without sharp crystalline peaks (Fig. 4A), demonstrating that this strategy consistently yields amorphous products over a wide compositional space. For the representative ten-element catalyst, ADF-STEM imaging and the corresponding EDS elemental maps further reveal a homogeneous spatial distribution of all constituent elements without obvious phase segregation (Fig. 4D), supporting the chemical uniformity of the resulting amorphous phase. As exemplified by the quinary $FeCoNiMoBO_x$ system, equimolar volumes of 0.2 M $Fe(NO_3)_3$, $Co(NO_3)_2$, $Ni(NO_3)_2$, and $(NH_4)_6Mo_7O_{24}$ solutions were first prepared and maintained at 4 °C, while a separately prepared 5 M $NaBH_4$ solution was refrigerated for 72 h before use. In an ice bath, the metal-salt solutions were combined, followed by rapid addition of twice the volume of the hydrolyzed $NaBH_4$ solution under vigorous stirring for 30 min; the resulting mixture was then kept at 4 °C for 18 h to suppress crystallization as completely as possible. We consider the use of highly concentrated hydrolyzed $NaBH_4$ to be critical, because its hydrolysis generates a strongly alkaline environment rich in boron–oxygen species and reducing borohydride complexes. Under these conditions, the boron–oxygen species can inhibit rapid metal migration while simultaneously coupling to the metal centers, and boron derived from the reducing species can be incorporated into the oxide matrix. These synergistic effects collectively favor the formation of homogeneous amorphous metal borate/oxide networks.

Having established the generality of this synthetic route, we next evaluated the OER performance of the synthesized compositions. Among them, the fully amorphous quinary $FeCoNiMoBO_x$ catalyst exhibited the best activity, delivering an overpotential of only 222 mV at 10 mA $cm^{-2}$, markedly lower than those of commercial $IrO_2$ and crystalline layered double hydroxides (LDHs) (Figs. 4B,C). We therefore selected $FeCoNiMoBO_x$, as it provides both optimal catalytic performance and a well-defined compositional platform for isolating structure–property relationships. Within this system, the boron content could be systematically tuned by varying the reaction time. Specifically, reactions carried out for 3, 6, and 18 h yielded three $FeCoNiMoBO_x$ samples with distinct boron loadings, denoted as Group-1, Group-2, and Group-3, respectively. ICP-OES analysis confirms that the experimentally obtained compositions are highly consistent with the intended designs, with boron contents of 5.87, 8.47, and 11.68 at.%, closely matching the nominal 6, 9, and 12 at.% levels (Supplementary Table S4). This well-controlled composition series therefore provides an appropriate experimental platform for testing the boron-dependent structural evolution predicted by ApolloX.

We then carried out multimodal structural characterization to benchmark the synthetic strategy and to test the predicted boron-driven amorphization trend. XRD results suggest that weak oxide-related peaks progressively broaden and diminish with increasing boron content (Fig. 5A). Synchrotron XRD further shows that, although all three samples lack sharp Bragg reflections (Supplementary Fig. S8), the diffraction features become progressively broader and weaker from Group-1 to Group-3, consistent with a gradual transition from partially crystalline to predominantly amorphous structures. The corresponding PDFs (Fig. 5B) likewise show the progressive disappearance of weak long-range periodicity with increasing boron incorporation.

Together, these observations indicate that higher boron contents are associated with a stronger tendency toward amorphization, in agreement with the theoretical predictions. Short-range PDF analysis (Fig. 5C) further suggests that this structural evolution is accompanied by disruption of M–O–M linkages, likely because increasing boron redistributes oxygen from the parent oxide network and promotes M–M interactions. At the same time, the remaining M–O coordination helps maintain structural homogeneity and suppress macroscopic phase separation.

The X-ray absorption measurements provide further support for this local structural evolution. XANES results indicate that the transition-metal elements retain broadly similar average oxidation states across the three groups (Supplementary Fig. S10). In the R-space EXAFS spectra at the Co K-edge (Supplementary Fig. S11), increasing boron content weakens the Co–O contribution while strengthening Co–M (M = metal) coordination, consistent with a progressive breakdown of oxide-like M–O–M connectivity and the emergence of stronger metal–metal associations.

Similar trends are also observed at the Ni and Fe K-edges. Wavelet-transform analysis of the Co K-edge EXAFS (Fig. 5E) shows a systematic shift of the [$\chi_{(k)}$, $\chi_{(R)}$] intensity centroid from Co–O–Co to Co–M scattering as the boron content increases, and the corresponding Co EXAFS fits in R space (Fig. 5D) reproduce the experimental features with low R-factors (Supplementary Table S5). Consistently, the short-range-order parameter $\alpha_{ij}$ quantitatively captures the same trend, namely reduced metal–oxygen connectivity and enhanced metal–metal association with increasing boron content (Supplementary Fig. S9). Taken together, the XRD, PDF, microscopy, and XAS results support a coherent structural picture in which increasing boron content progressively drives $FeCoNiMoBO_x$ away from oxide-like long-range order toward a more homogeneous amorphous network with reconfigured local coordination, broadly consistent with the structural evolution predicted by ApolloX.

## D. Amorphization strategy for multicomponent boroxide catalysts and functional validation in $FeCoNiMoBO_x$

To probe the boron-dependent structure–property relationships, we investigated three $FeCoNiMoBO_x$ compositions (Groups 1–3), whose metal ratios remained nearly constant. Electrochemical measurements show a clear activity trend: at high current density (100 mA $cm^{-2}$), the overpotentials decrease from 515 to 483 and 456 mV from Group-1 to Group-3 (Fig. S12), indicating that higher boron content is associated with improved OER performance under demanding operating conditions. To better understand this trend, we combined operando spectroscopy with electronic-structure analysis. *In situ* near-ambient-pressure X-ray photoelectron spectroscopy (NAP-XPS) shows that Co undergoes the most pronounced potential-dependent surface-state evolution in the multicomponent $FeCoNiMoBO_x$ matrix, whereas the Fe 2*p* and Ni 2*p* spectra change much less (Fig. S13). These observations identify Co as the most responsive redox-active surface center under OER conditions and motivates the use of Co-centered electronic descriptors. Within this framework, $\lambda_1$ and $\lambda_2$ are used as PDOS-derived initial-state descriptors [50] that quantify the degree of unoccupied Co 3*d* states and the number of electronically accessible active states for interfacial charge transfer, respectively (Supplementary Fig. S14). Both descriptors increase monotonically with boron content, consistent with the experimentally observed decrease in overpotential from Group-1 to Group-3. This agreement suggests that boron incorporation systematically modifies the Co-centered electronic structure in a manner favorable for OER, although it does not by itself identify the full operando active phase or isolate a single microscopic origin of activity.

We further assessed durability in an anion-exchange-membrane water electrolyzer (AEMWE). At 0.5 A $cm^{-2}$, all three $FeCoNiMoBO_x$ electrodes show stable operation, while Group-3 exhibits the lowest voltage degradation rate, 0.227 mV $h^{-1}$ over 800 h, substantially better than commercial $IrO_2$ under identical conditions (Fig. S12). Operando Raman spectroscopy of Group-3 likewise shows no detectable crystalline metal-oxide vibrational features across the applied potential range, while retaining a persistent -OH-related band near 773 $cm^{-1}$, indicating that the catalyst maintains a structurally robust disordered framework during operation.

Taken together, these results support a consistent structure–property interpretation for $FeCoNiMoBO_x$: increasing boron content correlates with changes in both local structure and electronic structure, including reduced M–O–M connectivity, enhanced Co–M/M–M near-neighbor correlations, and more favorable Co-centered initial-state electronic descriptors. These effects collectively align with the experimentally observed improvements in activity and durability and suggest that boron tunes the local catalytic environment in a cooperative way.

**Discussion**

Here we show that short-range-order-guided generative design can provide a practical route to compositionally complex amorphous oxides. By coupling ApolloX with boron-assisted synthesis, we identify a physically interpretable amorphization mechanism in which boron stabilizes $BO_3$-rich local environments, suppresses atomic diffusion, and promotes disorder while preserving functionally relevant metal–metal interactions. The agreement between theory and experiment in $FeCoNiMoBO_x$ suggests that this framework can move amorphous materials design beyond empirical trial-and-error toward structure-informed exploration.

At the same time, several limitations of the current framework should be noted. First, the PDM representation is designed to encode pair-level local chemical environments in a fixed-dimensional form, and therefore does not yet fully capture medium-range connectivity, angular correlations, or longer-range topological heterogeneity that can also shape amorphous functionality. A second limitation concerns the catalytic analysis: the present λ-descriptor framework is intentionally used as an initial-state electronic descriptor to rationalize composition-dependent OER trends, but it does not constitute a complete atomistic description of the reconstructed operando active phase or of the full multistep reaction pathway under working conditions. Future work should therefore integrate higher-order structural descriptors, closer experiment-theory feedback, and property-driven generative objectives to extend this strategy across broader amorphous material systems.

## Materials and Methods

### A. Pair-Density Matrix (PDM) descriptor

To encode SRO in the Cond-CDVAE framework both efficiently and without bias, we introduced the PDM as the primary structural descriptor. The PDM counts the number of near-neighbour atomic pairs for each element-element combination within a specified cutoff radius and arranges these counts in a K × K matrix, where K denotes the number of distinct element types. This representation maintains a fixed dimensionality regardless of composition while fully capturing local chemical environments. The PDM is formally defined in Equation (1):

$$PDM_{\alpha\beta}(r_c) = \sum_{i<j} \Theta\left(r_c - d_{ij}\right)\left[\delta_{si,\alpha,}\delta_{sj,\beta} + \delta_{si,\beta}\delta_{sj,\alpha}\right]$$

where $N$ is the total number of atoms in the simulation cell, $i,j \in \{1,...,N\}$, $s_i$ denotes the chemical species of atom i (e.g. Fe, Ni, B), and dij is the minimum-image interatomic distance under periodic boundary conditions. The cutoff radius is fixed to $r_c$ = 5.0 Å. $\Theta_{(x)}$ is the Heaviside step function ($\Theta_{(x)} = 1$ for $x > 0$, else 0) and $\delta_{a,b}$ is the Kronecker delta. This unnormalized PDM is closely related to the pair-probability form of the multicomponent Warren–Cowley/GM-SRO descriptors [43, 44]. Furthermore, similar pair-count descriptors are widely used as radial symmetry functions in modern machine-learning interatomic potentials[45], emphasizing the general applicability of this construction. Supplementary Text (Part A) provides the pseudocode for calculating the PDM along with a worked example to illustrate the procedure.

## B. Model details of Cond-CDVAE

The initial training set of 10,000 structures was generated by randomly occupying atomic sites in fixed-lattice structures. These configurations are then optimized using a fine-tuned DPA-2[29], a machine-learning potential capable of handling a wide range of elements. After optimization, each structure is labeled by its PDM.

We adapted the Cond-CDVAE model to generate structures conditioned on the PDM. In our scenario, the conditioning variables include both composition and PDM. The element type of each atom is represented by a categorical embedding vector. The composition is then calculated as a weighted average of the element type embedding vectors based on the number of atoms of each type. Meanwhile, the PDM, represented as a matrix of continuous variables, is flattened into a vector and normalized in the training set. The composition vector and PDM vector are concatenated to form the conditional vector. Other model hyperparameters are provided in Supplementary Table S3. Once trained, Cond-CDVAE[46] can generate new structures conditioned on the specified PDM, offering a versatile framework to explore diverse SRO configurations in high-entropy materials.

During the generation stage, the target composition and PDM should be provided. These conditioning variables, combined with a latent vector sampled from the latent space, are used by the lattice predictor to predict lattice parameters. Subsequently, the PGNNDec reconstructs a valid structure from random atomic coordinates using Langevin dynamics.

## C. Mean relative error of PDMs

To evaluate the reconstruction error distribution, diversity metrics (uniqueness and coverage), and novelty fraction, it is essential to define a standard for structural similarity.

A widely used criterion is the StructureMatcher function in the pymatgen library, which compares crystal structures by accounting for lattice parameters, atomic positions, species ordering, and symmetry operations within specified tolerances to determine whether two structures are crystallographically equivalent. However, matching becomes significantly more challenging as the number of atoms increases [46], and for multi-component alloys, the properties are not always dictated by the precise coordinates of individual atoms. In a 120-atom system, almost none of the structures are identified as equivalent by this method. Instead, we employ a parameter related to the local atomic environment—the mean relative error (MRE) of PDMs—to quantify structural similarity. The MRE is computed by evaluating the relative error of corresponding elements in two PDMs and then averaging over all elements. Because the relative error is asymmetric under the exchange of two structures, we define the MRE between two structures as the smaller of the two values obtained. Two structures A and B are considered equivalent if $\mathrm{MRE}(A,B) < \delta$, where $\delta$ is a predefined threshold.

## D. Finetune details of DPA-2

The structural complexity of the $B_{12}Mo_{12}Co_{12}Fe_{12}Ni_{12}O_{60}$ system poses significant challenges for structure optimization using first-principles methods. While MLPs have emerged as a promising approach to accelerate quantum mechanical calculations, enabling high-throughput, large-scale, and long-timescale atomic simulations at a significantly reduced computational cost, their accuracy and transferability critically depend on the quality of the training set. However, generating training data from scratch is both computationally expensive and labor-intensive, and MLPs often struggle to generalize beyond the scope of their training data. To address these limitations, we fine-tune the pre-trained universal DPA-2 model[29], which employs a universal descriptor trained on a diverse dataset computed at different levels of theory and can be further adapted to specific chemical systems of interest. The DPA-2 descriptor consists of a repinit layer to embed the local atomic environment with a cutoff radius of 6.0 Å; the atomic representation is subsequently updated using
six message-passing repformer layers. Specifically, we refine the DPA-2 model using trajectories obtained from ab-initio MD simulations consisting of 2789 frames under a random branch, whose fitting network consists of three hidden layers with 240 neurons each. After 80000 training steps, the resulting model achieves a root mean square error (RMSE) of 7.76 meV/atom in energy, 139.61 meV/ Å in force, and 0.068 meV/atom in virial, demonstrating its efficacy in accurately describing the targeted system (detailed in Supplementary Fig. S4).

## E. Warren-Cowley SRO parameter

To describe the local environment of multi-element materials, the Warren–Cowley SRO parameter [42] is often defined as:

$$\alpha_{ij}(r) = 1 - \frac{P_{ij}(r)}{c_j}$$

where i, j denote atomic species; $P_{ij}$(r) is the conditional probability of finding an atom of type j at distance r (or within a specific neighbor shell) around a central atom of type i; and cj is the overall molar (or atomic) fraction of species j in the system. If $\alpha_{ij}$(r) = 0, it indicates that the distribution of species j around species i matches the global composition; if $\alpha_{ij}(r)$ >0, it implies that species j is deficient (repelled) around species i; if $\alpha_{ij}$(r) <0, it implies that species j is enriched (aggregated) around species i. In an ideal random distribution, $\alpha_{ij}$(r) = 0 for any i, j, and r. When the system exhibits short-range order, $\alpha_{ij}$(r) deviates from 0.

## F. DFT and MD calculations

In DFT calculations, the Perdew Burke-Ernzerhof generalized gradient approximation and projector-augmented wave (PAW) pseudopotentials implemented in VASP [47–49] code were employed. For Fe, Co, Ni, Mo, and B, the number of valence electrons of 8, 9, 10, 6, and 3 were explicitly considered, respectively. The corresponding valence electron configurations are Fe ($3d^6$ $4s^2$), Co ($3d^7$ $4s^2$), Ni ($3d^8$ $4s^2$), Mo ($4d^4$ $5s^2$), and B ($2s^2$ $2p^1$), which are in agreement with the ZVAL values obtained from our pseudopotential setup. Core electrons were treated using PAW potentials. A plane-wave cutoff energy of 650 eV and Γ-point-only Brillouin-zone sampling were used for structural optimizations, with convergence criteria of 1 meV atom$^{-1}$ in total energy and 10 meV Å$^{-1}$ in residual forces.

To refine the generated structures, we fine-tune the pretrained DPA-2 model [29] with selected MD trajectories frames to achieve both accuracy and efficiency in searching for local minima. Geometry optimization with L-BFGS method implemented was employed to relax each candidate structure into a DPA-2-based local minima. The fitness function for these simulations was the free energy at T = 0 K, which reduces to enthalpy, ensuring the physical relevance of the predicted configurations. Although local optimization increases computational cost, it reduces noise in the energy landscape, improves configuration comparability and produces optimized structures for subsequent analysis.

## Acknowledgments

This work was supported by the National Natural Science Foundation of China (Grants No. T2225013, No. 12034009, No. 12174142, No. 42272041, No. 22372004, No. 22405048), National Key Research and Development Program of China (Grants No. 2022YFA1402304, No. 2024YFA1509500), Beijing Natural Science Foundation No. Z240027, Program for Jilin University Science and Technology Innovative Research Team (2021TD–05), Program for Jilin University Computational Interdisciplinary Innovative Platform. Part of the calculation was performed in the high-performance computing center of Jilin University, Postdoctoral Fellowship Program of CPSF under Grant Number GZC20252243. W.-L.L. and H.Z. thank the KAUST Supercomputing Laboratory for providing computational resources on the Shaheen III supercomputer through project k10175. We thank Prof. Zipeng Zhao for carrying out the catalytic-stability measurements. We acknowledge Prof. Xueqiang Zhang for the in-situ NAP-XPS studies conducted at BL02B Beamline of Shanghai Synchrotron Radiation Facility. We thank Prof. Jihan Zhou for the Electron Microscopy Laboratory at Peking University for the use of aberration-corrected electron microscope. We are also grateful to Yudong Wang for his expert assistance with X-ray diffraction experiments performed at beamline BL17UM of the Shanghai Synchrotron Radiation Facility.

## Author contributions

M.L., Y.W., and W.-L.L. supervised the research and project. H.L. designed the workflow of ApolloX. C.L., Z.L., H.Z., S.Z., C.P., Q.J., Y.W. and J.L. performed the experimental component, including synthesis, characterization, and performance testing. Y.G., X.L., J.L., and Y.D. contributed to code development and model design, while Y.C. and J.W. analyzed and interpreted the data, conducted performance evaluations, and performed model comparisons. G.L. and Y.L. carried out catalytic property calculations. R.W., Z.W., and Z.Z. contributed to training the machine learning potential. Additionally, H.L., C.L., M.L., Y.W., W.-L.L., X.L., C.M., J.L. and Z.X. contributed to the writing of the manuscript.

## Competing interests

The authors declare no competing interests.

## Data and materials availability

Supplementary information The online version contains supplementary material available at XXX. The authors declare that the main data supporting the findings of this study are contained within the paper and its associated Supplementary Information. The ApolloX source code is available on GitHub (https://github.com/FNC001/ApolloX)

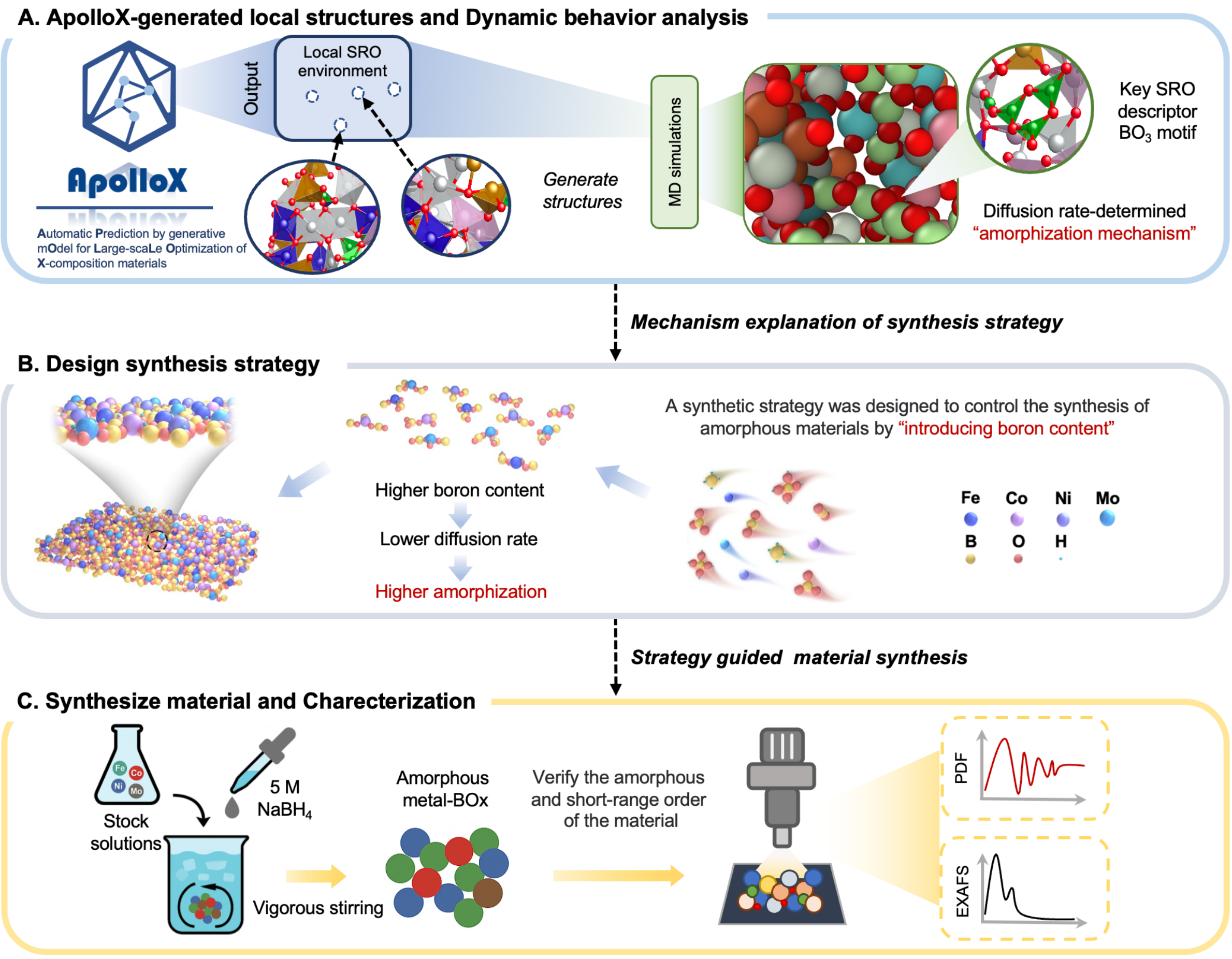

**FIG. 1 | ApolloX-guided design of boron-assisted amorphous multi-metal BOx catalysts. (A)** ApolloX workflow for generating statistically representative local SRO environments in multi-metal $BO_x$ compositions, enabling direct visualization of metal–oxygen and boron–oxygen coordination motifs. Dynamic behavior analysis by *ab initio* molecular dynamics (MD) based on ApolloX-derived structures, revealing the emergence of $BO_3$ motifs that act as diffusion-slowing units and promote amorphization. **(B)** Translation of these insights into a boron-content-programmable synthetic strategy within a controlled compositional window, in which the boron content is systematically tuned while the overall compositional regime is controlled; increasing boron content strengthens $BO_3$-rich networks and enhances the tendency toward amorphous structures. **(C)** Experimental realization and validation of the designed multi-metal $BO_x$ catalysts, including amorphous nanoparticle morphology, diffuse electron-diffraction halos.

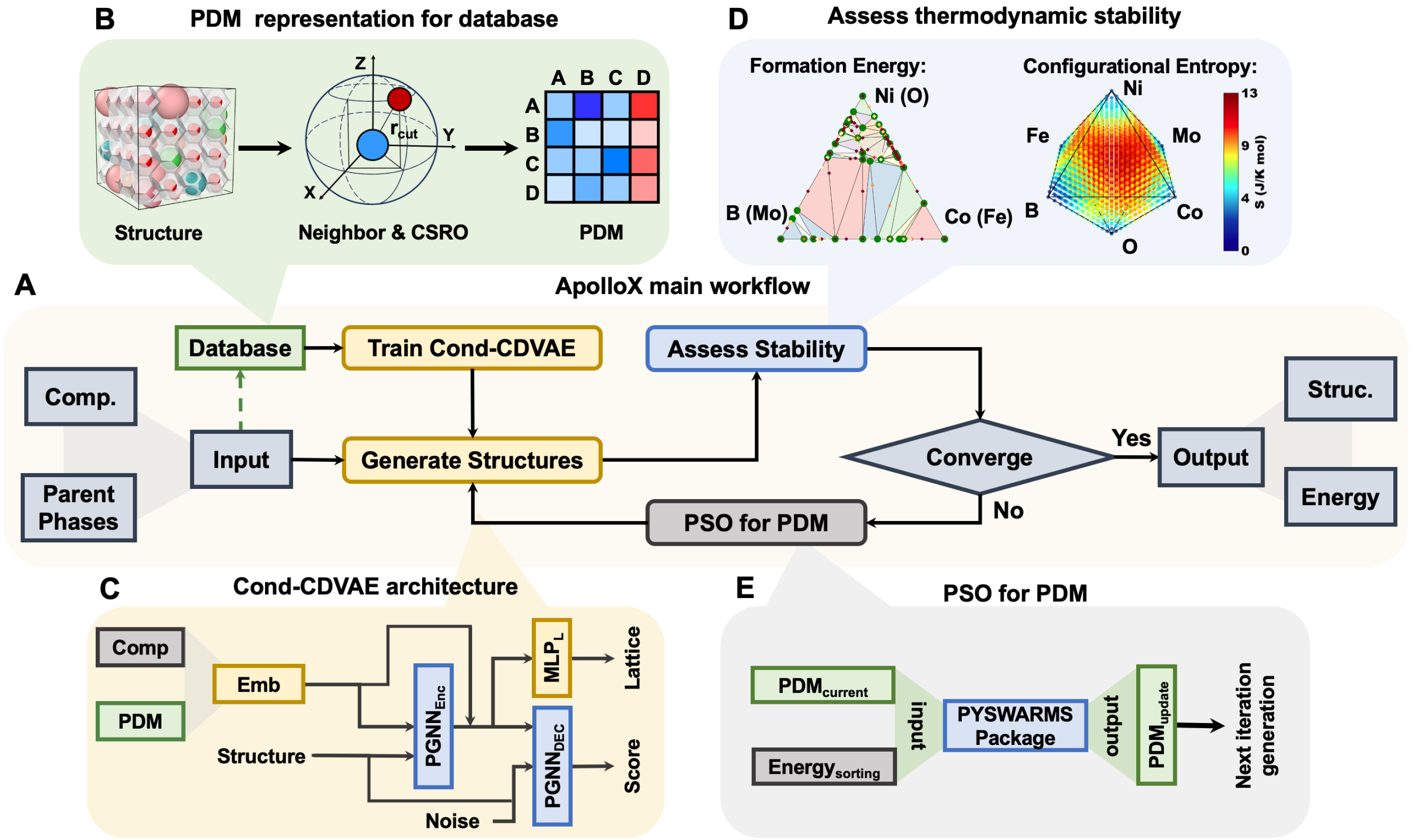

**FIG. 2 | Overview workflow of ApolloX. (A)** Main workflow of ApolloX for structural modeling, including the model's iterative generation process. **(B)** Structures are represented using the Pair Density Matrix (PDM), and a dataset is constructed by mapping each structure to its corresponding PDM. **(C)** The model architecture of the Cond-CDVAE. An embedding network (EMB) integrates a structure's composition (Comp.) and PDM into a unified conditioning vector. The graph neural network encoder (PGNNEnc) encodes the structure alongside its conditioning vector into a latent vector, capturing its underlying features. Subsequently, a multi-layer perceptron MLPL predicts the lattice parameters based on the latent vector. And the graph neural network decoder (PGNNDec) outputs a score function for denoising and reconstruction of a plausible structure. **(D)** Enthalpy stability is evaluated using the convex hull method, and entropy stability is assessed based on configurational entropy. **(E)** PSO is employed to optimize the PDM, using the PYSWARMS package.

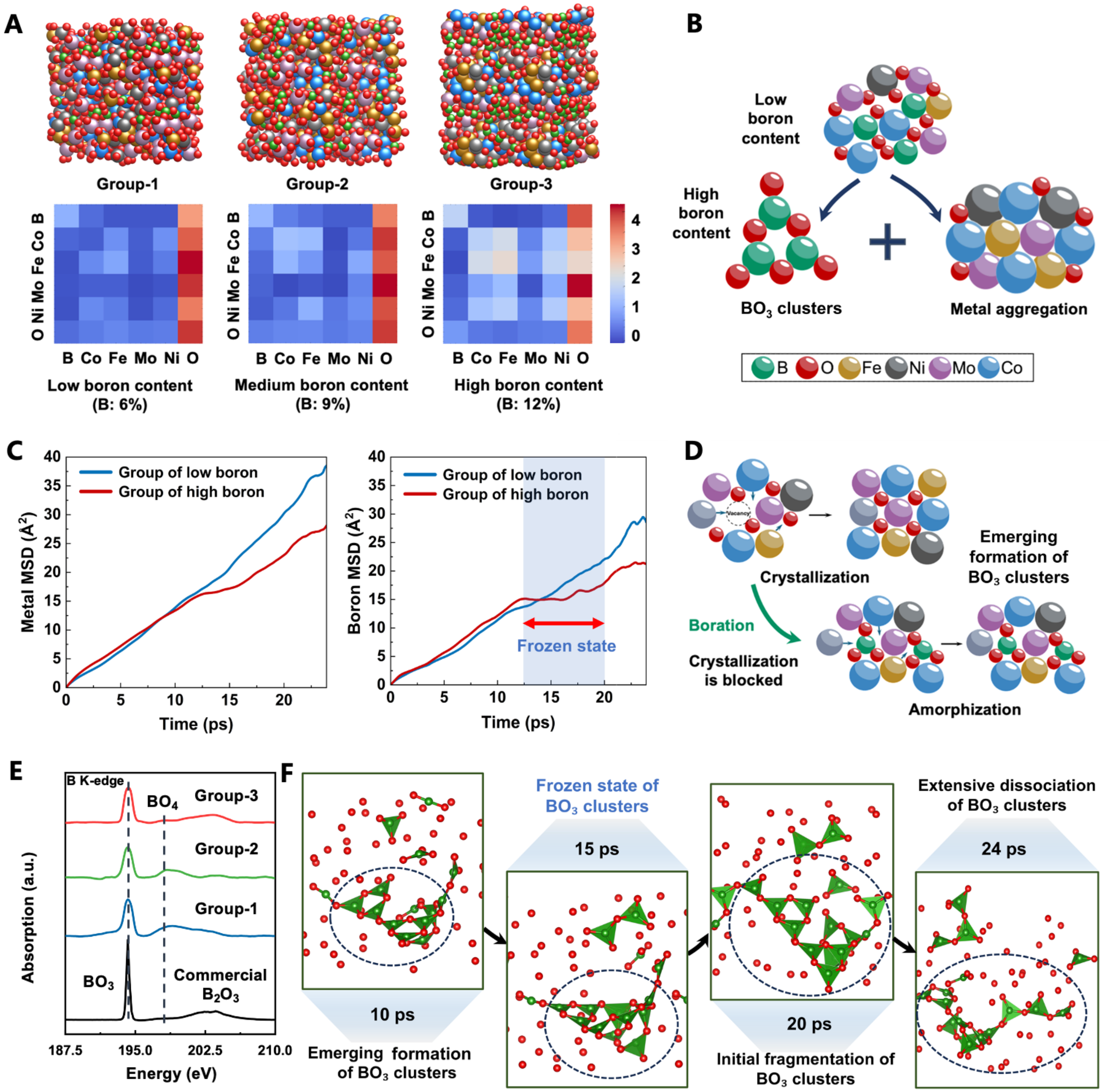

**FIG. 3 | Short-range order evolution and diffusion-limited amorphization mechanism. (A)** Nearest-neighbor statistics heatmaps for three boron contents. **(B)** Higher boron promotes short-range B–O ordering and metal clustering. **(C)** MD-derived diffusion coefficients show boron diffuses more slowly than metals (300–2050 K, 24 ps). **(D)** Enriched $BO_3$ motifs reduce mobility, hinder rearrangement, and favor amorphization. **(E)** B K-edge XANES of Group-1/2/3 vs. $B_2O_3$: $BO_3$ (~194 eV) and $BO_4$ (~198 eV); increasing $BO_3$ indicates trigonal boron enrichment. **(F)** Time-resolved snapshots show $BO_3$ clusters evolve into a percolating network (structural arrest) before partial breakup at high T.

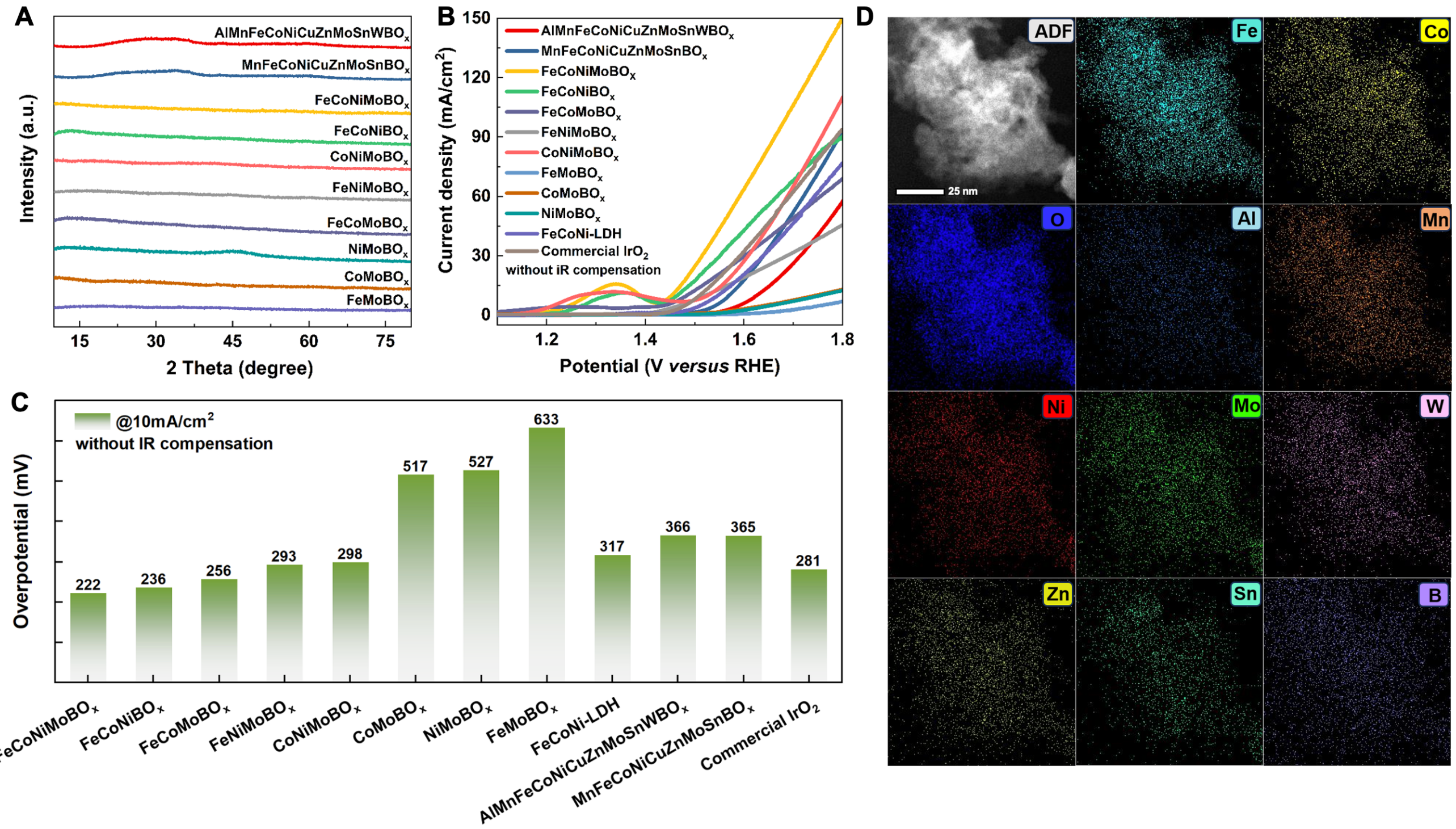

**FIG. 4 | Universality of the synthetic strategy. (A)** X-ray diffraction (XRD) patterns of the synthesized boron-containing multimetal oxides, from medium- to high-entropy compositions, all exhibiting broad diffuse features characteristic of amorphous structures. **(B)** Linear sweep voltammetry (LSV) curves for the oxygen evolution reaction (OER) measured in 1.0 M KOH without *iR* compensation. **(C)** Corresponding overpotentials required to reach a current density of 10 mA $cm^{-2}$. **(D)** ADF-STEM image and corresponding EDS elemental maps of the ten-element high-entropy amorphous catalyst, $AlMnFeCoNiCuZnMoSnWBO_x$, showing the homogeneous spatial distribution of the constituent elements.

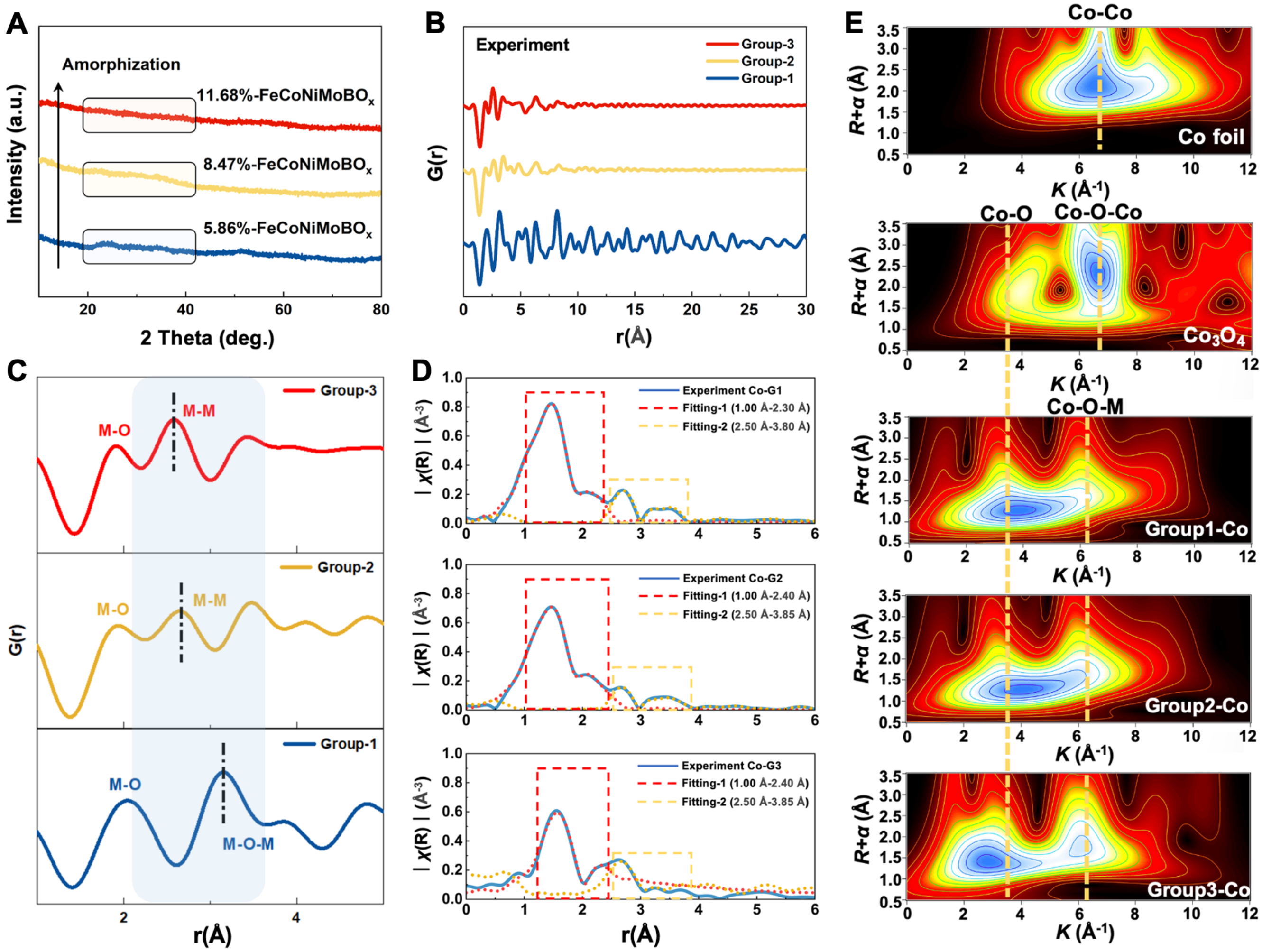

**FIG. 5 | Structural characterization of FeCoNiMoBOx samples with varying boron contents. (A)** X-ray diffraction (XRD) patterns of Group-1, Group-2, and Group-3 samples, showing the disappearance of long-range diffraction peaks and progressive amorphization with increasing boron content. **(B)** Experimental pair distribution functions (G(r)) revealing changes in local atomic ordering. **(C)** Short-range G(r) curves highlighting the relative intensities of M–O, M–M, and M–O–M correlations for each group. **(D)** Fourier-transformed EXAFS fitting curves in the R-space EXAFS spectra of the Co K-edge, showing a decrease in Co–O coordination and an increase in Co–M interactions as boron content increases. **(E)** Wavelet-transform EXAFS (WT-EXAFS) contour plots of Co foil, $Co_3O_4$, and the three sample groups, where the shift from Co–O–Co to Co–M scattering paths with higher boron levels is evident.